\documentclass[aps,pre,twocolumn,showpacs,floatfix]{revtex4-1}
\usepackage{graphicx}
\usepackage[version=3]{mhchem}
\usepackage{amsmath}
\usepackage{sidecap}
\usepackage{floatrow}
\usepackage{epstopdf}
\usepackage{texlogos}
\usepackage{wasysym}
\def\beq{\begin{equation}}
\def\eeq{\end{equation}}
\def\bea{\begin{eqnarray}}
\def\eea{\end{eqnarray}}
\begin{document}
\makeatletter
\title{Enhancement of Network Synchronizability via Two Oscillatory System}
\author{Harpartap Singh}
\email{harpartap@mail.saitama-u.ac.jp}
\affiliation{Division of Strategic Research and Development, Graduate School of Science 
and Engineering, Saitama University, Shimo-okubo 255, Sakura-ku, Saitama 338-8570, Japan}
\date{\today}
\begin{abstract}
The loss of synchronizability at large coupling strength is of major concern especially 
in the fields of secure communication and complex systems. Because theoretically, the 
coupling mode that can surely stabilize the chaotic/hyperchaotic synchronized state is vector coupling 
(using all the coordinates) which is in contrast to the practical demand of information exchange using lesser 
number of coordinates (commonly via a single coordinate).  In the present work, we 
propose that if the node dynamics are given by a pair of oscillators (say, {\it two oscillatory system} TOS) 
rather than by a conventional way of single oscillator (say, {\it single oscillatory system} SOS), 
then the information exchange via a single coordinate could be sufficient to 
stabilize the chaotic/hyperchaotic synchronization manifold at large coupling strength. The frameworks of drive-response system and 
Master Stability Function (MSF) have been used to study the TOS effect by varying TOS parameters with and without 
feedback (feedback means quorum sensing conditions). The TOS effect has been found numerically both in 
the chaotic (R{\"o}ssler, Chua and Lorenz) and hyperchaotic (electrical circuit) systems. 
However, since threshold also increases as a side effect of TOS, the extent of $\beta$ enhancement 
depends on the choice of oscillator model like larger for R{\"o}ssler, intermediate for Chua and smaller 
for Lorenz.
 \end{abstract}
\maketitle
\section{Introduction}

Synchronization is not always an obvious emergent behavior of the interacting 
dynamical systems even in the case when they are identical limit cycle oscillators. For an example, 
$N$ identical R{\"o}ssler oscillators (periodic/chaotic) coupled via a single coordinate (scalar coupling) on a ring network 
in the nearest neighborhood configuration exhibit desynchronization at large coupling strength~\cite{heagyprl2,pecorapre}. 
This desynchronization behavior (known as Short Wavelength Bifurcation SWB~\cite{heagyprl2}) arises 
because the minimum coupling strength (threshold) requires for the synchronization 
increases with the increase in number of oscillators whereas the synchronization manifold in case of scalar coupling mode, 
losses its stability (riddle basin behavior~\cite{heagyprl1,pecorachaos}) as the coupling strength ($\gamma$) augments towards a 
critical value (overload-tolerance). Thus, the small values of overload-tolerance (say $\gamma_{dsyn}$) limit the size of a 
synchronizable network, e.g. the synchronized state of a ring network having $x_1$-coupled chaotic R{\"o}ssler becomes 
unstable with the increase in $N$ from $18$ to $19$ at $\gamma_{dsyn}=$1.5. Moreover, since the mechanism which could increment 
$\gamma_{dsyn}$ has not been developed so far, the network modification methods are generally used to 
tackle the issue of strong coupling such as (1) adding the additional edges 
between the nodes deterministically (Pristine World) or/and stochastically (Small World) ~\cite{barahona}, (2) 
modifying the ring network topology to a synchronizable topology like a unidirectional tree network or a star network 
(a hub of Scale Free network)~\cite{kurthpre,kurthchaos,kurthphyrep1}. Practically these 
modifications could be considered as the distribution of overload among the nodes and theoretically these alterations imply the 
minimization of eigen-ratio ($R$) of the Laplacian/coupling matrix (largest eigenvalue to the smallest nonzero eigenvalue).  
The concept of $R$ minimization comes from the theory of  Master Stability Function (MSF)~\cite{pecoraprl} which says that a 
complex network of size $N$ is synchronizable if $R<\beta$ where $\beta=\gamma_{dsyn}/\gamma_{syn}$ ($\gamma_{syn}$ 
means threshold)~\cite{barahona}. Therefore, to study the chaotic complete synchronization behavior 
in the complex systems, finding the different pathways that could reduce $R$ remain a primary goal for the 
researchers~\cite{barahona,kurthpre,kurthchaos,kurthphyrep1,pecoraprl,boccalettiaphyrep,lin,Nishikawaprl,boccalettiaprl,donetti,Nishikawapre,lu,buldu,Boccalettipre,amritkar}. 
However, modification (1) increases the coupling cost of synchronization whereas the modification (2) 
decreases the robustness by increasing the centralization in a network. 

In summary, the primary question which has not been answered yet is ``how to achieve the 
finitely large overload-tolerance in case of scalar coupling as similar to the vector coupling scenario (wherein all coordinates are used), i.e. 
{\it maximization of $\gamma_{dsyn}$ via scalar coupling}''? The importance of this question lies on the fact that 
getting stability by using all the coordinates is neither useful (as in case of secure communication) nor realistic (as in case of complex system).
Moreover, this theoretical problem also appears in many different forms such as: is it possible to stabilize the chaotic/hyperchaotic synchronization manifold? or 
is it possible to make a large non-centralized chaotic/hyperchaotic synchronizable network? or 
is it possible to surely stabilize the chaotic/hyperchaotic sub system under local/global parameter fluctuations?, etc. 
The answer to this question could be considered as a possible solution to the real problems such as 
overload failure in the real networks like Internet system and power grid system~\cite{holme,motter},  
the stability issue of chaotic/hyperchaotic transmitter-receiver system in the field of secure communication~\cite{pecorachaos,pecora,pecorapra}, etc. 
It should be noted that since large coupling strength means infinite coupling ($\gamma=\infty$) 
in drive-response system~\cite{pecorachaos,pecora,pecorapra}, 
MSF complements the drive-response formulation by explicitly incorporating  
$\gamma$ dependence which results into an elegant relation between the node property ($\beta$) of a 
 just two node system (coupled bidirectionally) with the structural property ($R$) of an arbitrary network having $N$ nodes 
 (provided coupling matrix has zero row sum). In other words, if a system is stable in the drive-response framework then at the 
 large coupling strength MSF shows negative values and vice-versa (discussed later).

\begin{figure}
\includegraphics[height=5cm,width=8cm,angle=0]{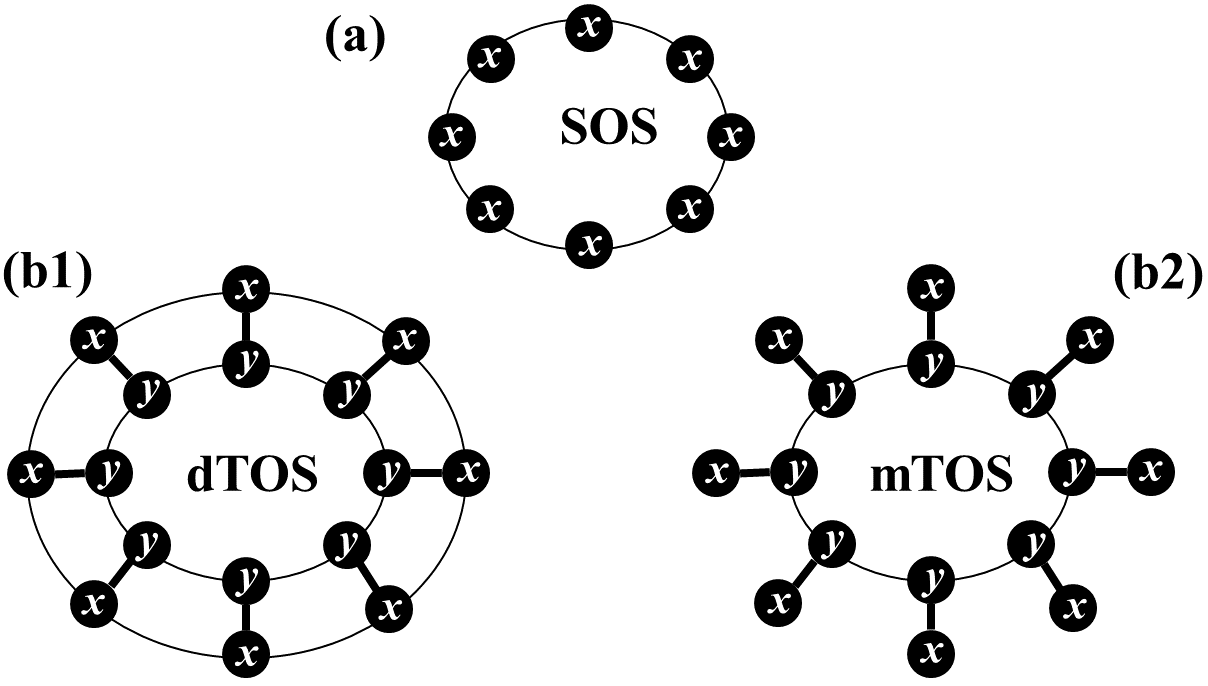}
\caption{Schematics of a ring network (nearest neighborhood configuration bidirectionally coupled) 
for two scenarios: each node behaves as (a) conventional {\it single oscillatory system} (SOS), (b) 
proposed {\it two oscillatory system} TOS. The subplots (b1) and (b2) depict the implementation of TOS, i.e. 
by using both the oscillators (dTOS) and one of the oscillator (mTOS), respectively, where `d' means `di' and 
`m' means `mono'. Oscillators $x$ and $y$ are coupled via all the coordinates (vector coupling) whereas oscillators 
$x$ and $x$ ($y$ and $y$) are coupled via a single coordinate (scalar coupling).} 
\label{1}
\end{figure}

In the present work, an attempt has been made to answer the primary question of maximization of $\gamma_{dsyn}$ 
via scalar coupling in case of linear interactions under small perturbations and global parameter fluctuations 
(each node experiences same fluctuations, i.e. identical node scenario). We argue 
that $\gamma_{dsyn}$ could be maximized, if the node dynamics are given by a pair of oscillators 
(say, {\it two oscillatory system} TOS) rather than by a conventional way of single oscillator (say, {\it single oscillatory system} SOS) 
as schematic shown in Fig.~\ref{1} on a ring network (i.e. a basic non-centralized network). 
In addition, since the stabilization of synchronization manifold happens only due to the emergence of 
dissipative factors via TOS (explained later by drive-response framework), 
to maximize $\gamma_{dsyn}$ the following two conditions should be met: 1. TOS should be in mTOS 
configuration as dTOS configuration behaves same as SOS (discussed later), 2. only that coordinate can be employed 
whose self dissipation could stabilize the unstable fixed point by adding its linear dissipative term in the autonomous system (discussed later). 
According to condition 2, all the three coordinates of Chua~\cite{chua} and Lorenz~\cite{lorenz} oscillators fulfill this criteria whereas only 
two out of the three coordinates of R{\"o}ssler~\cite{rossler} (i.e. $x_1$ and $x_2$) and also only two out of the 
four coordinates of piecewise electronic hyperchaotic system~\cite{pecorahyper} (i.e. $x_1$ and $x_2$),  satisfy this condition. 
However, there also exist a possibility when none of the coordinates of an oscillator model like hyperchaotic R{\"o}ssler~\cite{rosslerhyper}, 
could meet the condition and hence can not be employed as the oscillatory dynamics for TOS. 
Furthermore, since threshold increases with the increase in number of oscillators, $\gamma_{syn}$ 
also increases as the side effect of mTOS along with the incrementation of $\gamma_{dsyn}$. This makes 
$\beta$ enhancement dependent on the choice of oscillator model like larger for R{\"o}ssler, intermediate for 
Chua and smaller for Lorenz.

The paper is organized as follows. In Sec. II, the stability issue of scalar coupling and 
the theory of TOS are given in terms of MSF and drive-response frameworks along with 
the equations of the employed oscillator models. The stabilization via TOS for
the scenarios of chaotic (R{\"o}ssler, Chua and Lorenz) as well as hyperchaotic (electrical circuit) oscillators 
are presented in Sec. III. Finally, the paper is concluded in Sec. IV.

\begin{figure}
\includegraphics[height=2cm,width=5cm,angle=0]{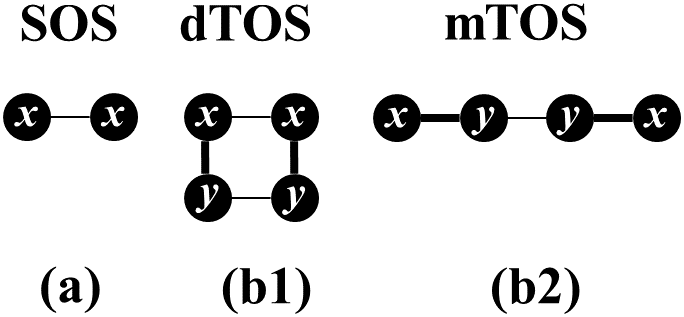}
\caption{Schematics of a two node network (unidirectional/bidirectional) wherein oscillators
$x$ and $y$ are coupled via all the coordinates (vector coupling) whereas oscillators 
$x$ and $x$ ($y$ and $y$) are coupled via a single coordinate (scalar coupling).} 
\label{2}
\end{figure}

\section{Theory}
Under small perturbations (linear analysis), Master Stability Function (MSF) serves as the necessary and sufficient 
condition to judge the basin stability at large coupling strength. This is analogous to the drive-response system 
wherein the conditional Lyapunov exponents surely determine the stability of a sub-system (response) by running the 
system from the different initial conditions. Therefore, the sub-Jacobian (Jacobian of response) method has been 
employed to investigate the reason behind the emergence of TOS effect.

\subsection{MSF}
To understand the formulism of MSF~\cite{pecoraprl}, consider the equations of motion for a 
complex network having $N$ identical nodes for the conventional {\it single oscillatory system} (SOS):\newline
$\dot{x}^i=F(x^i)+\gamma\sum\limits_{j=1}^{N}G_{ij}H(x^j)$\hspace*{4.0cm}(1)\newline
Where $x^i$ is a $m$-dimensional vector of $i^{th}$ node whose autonomous behavior is described 
by $F(x^i)$ ($R^m\rightarrow R^m$). Here, $H(x)$ ($R^m\rightarrow R^m$) represents $m\times m$ matrix (linear coupling function) 
which gives the information about the coordinates of $x$ involved in the coupling, i.e. its 
all other elements are zero except a diagonal element corresponding to the employed coordinate (for more details see ref.~\cite{pecoraprl}). 
In Eq. 1, $G$ is a $N\times N$ coupling matrix (captures network's architect) which could be symmetric or not but it must be real with zero 
row sum~\cite{kurthphyrep1} so that synchronous state ($x^{i}=s$, i=1,..,N; $\dot{s}=F(s)$) could become 
a solution of Eq. 1. For a symmetric case (Laplacian), $G_{ii}=-g_i$ ($g_i$ is the degree/connections of $i^{th}$ node) 
and $G_{ij}=1$ if $i^{th}$ node is connected to $j^{th}$ node otherwise $G_{ij}=0$. 
The parameter $\gamma$($>0$) represents the coupling strength. 

The $N$ block diagonalized linear variational equations for Eq. 1 that depict the stability of the synchronous state, i.e. 
evolution of perturbation ($\eta$), are~\cite{pecoraprl}:\newline
$\dot{\eta}^k=\large(DF(s)-\gamma \delta_kDH(s))\eta^k$\hspace*{3.9cm}(2)\newline 
Here $DF(s)$ and $DH(s)$ are the Jacobian matrices evaluated at $\dot{s}=F(s)$. 
In Eq. 2, $\delta_k$ ($k$=1,2,..,N) show $N$ non-negative real eigenvalues of $G$ (symmetric), 
such that $0=\delta_1\le$$\delta_2\le$..$\le\delta_N$. So, the structural property becomes $R=\delta_{N}/\delta_{2}$, 
as $\delta_2$ is minimum non-zero eigenvalue ($\delta_{min}$) and $\delta_N$ is maximum eigenvalue ($\delta_{max}$). 
Now to check the network's synchronizability, we need to evaluate node property $\beta$ from Eq. 2.  
For this, we proceed as follows. 

Corresponding to the $N$ eigenvalues, there are $N$ maximum Lyapunov exponents, $\lambda( \gamma\delta_k)$, wherein  
$\lambda(\gamma\delta_1)$ ($\lambda(0)$ or $\lambda_0$) describes the state of the synchronized regime as
$\lambda_0=0$ implies periodic and $\lambda_0>0$ means chaotic. The linear stability of this synchronized state 
is decided by the remaining $N-1$ maximum Lyapunov exponents (transverse), i.e. 
the given synchronous state is stable if $\lambda(\gamma\delta_k)<0$ ($k>1$), and these $N-1$ exponents 
can be found from a single variational equation by using scaling relation ($\gamma\delta_k$)~\cite{pecorapre}. This means 
that at fixed $\gamma$, the variational equation in the presence of $\delta_{max}$ may experience stronger coupling and 
hence its stabilization or destabilization may occur earlier than the other modes. In other words, 
$\lambda(\gamma\delta_{max})$ plays a vital role in the stability condition wherein 
all the modes should be simultaneously stabilized or one could say that 
$\lambda(\gamma\delta_{max})$ can solely provides the stability of a network 
which is the case and hence it is called MSF for given $DF(s)$ and $DH(s)$. This is because, as long as 
$G$ is diagonalizable its eigenvalues corresponding to different network topologies as well as sizes can be found from any
topology using scaling relation (for more details see ref.~\cite{pecorachaos,pecoraprl}). Therefore, MSF of two bidirectionally 
coupled nodes, i.e. $\lambda(2\gamma)$, yields same $\beta$ (=$\gamma_{dsyn}$/$\gamma_{syn}$) as an arbitrary network having $N$ nodes does,i.e. $\beta$ (=$\sigma_{dsyn}$/$\sigma_{syn}$) where $\sigma=\gamma\delta_{max}$. Now we can define 
$\gamma_{syn}$ and $\gamma_{dsyn}$ (similarly $\sigma_{syn}$, $\sigma_{dsyn}$), i.e. 
$2\gamma_{syn}$ is the minimum coupling strength at which $\lambda(2\gamma)$=$0$ or 
$\lambda(2\gamma)\rightarrow0^+$ (emergence of synchrony) and $2\gamma_{dsyn}$ is the maximum coupling strength at 
which $\lambda(2\gamma)$=$0$ or $\lambda(2\gamma)\rightarrow0^-$ (emergence of desynchrony). 

\subsubsection{Problem of scalar coupling}
 In contrast to the vector coupling scenario ($\lambda(\sigma\rightarrow\infty)<0$), 
in case of scalar coupling wherein $\lambda(\sigma)>0$ as $\sigma>\sigma_{dsyn}$ ($\sigma_{dsyn}$ is small), 
it becomes a challenge to synchronize a ring network topology (Fig.~\ref{1}(a)) because its 
$R$ (i.e. $1/ sin^2(\pi/N)$) grows faster with the increase in $N$ than any other topology, e.g. for a star network $R=N$. 
Therefore, the previous works on the enhancement of network's synchronizability using 
network modification methods could be considered as the work on the minimization of $R=1/ sin^2(\pi/N)$. 

\subsection{Drive-Response}
To understand the framework of drive-response~\cite{pecora,pecorapra}, consider the equations of motion 
for two unidirectionally coupled identical nodes for the conventional {\it single oscillatory system} SOS (Fig.~\ref{2}(a)):\newline
$\dot{x}^1=F(x^1)$\hspace*{4.0cm}\newline
$\dot{x}^2=F(x^2)+\gamma\Gamma(x^1-x^2)$\hspace*{4.0cm}(3)\newline
Where $\Gamma$ depict same information as $H(x)$ does in Eq. 1, e.g.
for $m=3$, i.e. $x$=($x_1$,$x_2$,$x_3$): 
$\Gamma$=diag(1,0,0)$_{3\times 3}$ or diag(0,1,0)$_{3\times 3}$ or diag(0,0,1)$_{3\times 3}$ (scalar coupling). 
Now we choose $x_1$ coordinate for the interactions and assume that the coupling term vanishes, i.e. 
$x^2_1\rightarrow x^1_1$, as $\gamma\rightarrow\infty$ which is only possible if $\dot{x}_1^2=\dot{x}_1^1$. 
This means that $x^2$ node instead of generating its own $x_1^2$ signal must use $x_1^1$ signal from 
$x^1$ node in order to ensure $x^2_1= x^1_1$ ($\gamma=\infty$) for all the time, i.e. even when $x^2_j\neq x^1_j$ ($j=2,..,m$). 
Hence, Eq. 3 becomes $\dot{x}^2=F(x^2,x_1^1)$ and $\dot{x}^2$=($\dot{x}_2^2$,..,$\dot{x}_m^2$). In this scenario, 
Eq. 3 depicts $x_1$-driving system wherein $x^1$ is drive and $x^2$ is a response sub system. The 
maximum conditional Lyapunov exponent ($\lambda^c$) of sub system (sub Jacobian) 
decides the stability of synchronization manifold, i.e. $\lambda^c<0$ implies stabilization (for more details see ref.~\cite{pecorachaos}). 
Thus, this becomes analogous to the MSF behavior of two bidirectionally $x_1$-coupled nodes wherein 
the condition of stabilized synchronization manifold is $\lambda(2\gamma)<0$. Therefore, by using scaling relation 
one could also relate with the $N$ nodes scenario, i.e. $\lambda(\gamma\delta_{max})<0$ (as discussed above).

\subsubsection{Problem of scalar coupling}
Since the practical version of drive-response system is transmitter-receiver system in the field of secure communication, 
the information transfer via a single coordinate remains a primary concern. But in case of SOS, only 
the specific coordinates of $x$ can be used as a drive (depending upon the choice of oscillator)
which sometimes can not provide the stabilization against the parameter fluctuation as the scenario of 
$x_2$-driving chaotic R{\"o}ssler (shown in ref~\cite{pecorapra}). Secondly, to stabilize 
a hyperchaotic drive-response usually a scalar signal is generated by using BK method~\cite{peng} which requires many parameter ($2m$).  
Hence, this necessitates some simpler mechanism which could work for both chaotic as well as hyperchaotic scenarios.

\subsection{TOS}
In contrast to SOS ($\dot{x}=F(x)$), the proposed {\it two oscillatory system} (TOS) is made up of two bidirectionally coupled oscillators (identical/nonidentical) 
by employing their all coordinates (vector coupling) so that the stable synchronization behavior (complete/generalized) could be ensured. 
Here, one should not confuse the used terminology of two oscillators with two nodes, i.e. 
in TOS scenario, the dynamics of a single node are generated by a pair of oscillators as:\newline
$\dot{x}=F(x,\mu)+\gamma_1\Gamma_{v}(y -x )$\newline
$\dot{y}=F(y,\mu')+\gamma_2\Gamma_{v}(x -y )$\newline
Similar to $x$, $y$ is also a $m$-dimensional vector whose autonomous behavior is described 
by same function $F(y)$ ($R^m\rightarrow R^m$) where $\mu$ and $\mu'$ are 
the intrinsic parameters of the models. The parameters $\gamma_1$ and $\gamma_2$ represent the 
intra-coupling constants of TOS and $\Gamma_{v}$ shows the vector coupling scenario between $x$ and $y$, i.e. 
$\Gamma_{v}$$=$diag(1,..,1)$_{m\times m}$. For the scenario of 
$\gamma_2$=$\theta\gamma_1$ (feedback), TOS could be considered as a two oscillatory representation of the 
quorum sensing network of $x$ oscillators interacting with each other via a $y$ oscillator (medium)~\cite{singh}. The parameter 
$\theta$ represents the population density of $x$ at each node of a network whereas $y$
may have periodic/chaotic dynamics, i.e. other than the conventional steady state behavior used in quorum sensing~\cite{monte,russo,singh}. 

Further, the nodes having TOS dynamics may interact with each other in two ways, i.e. (i) each node uses either 
its $x$ or $y$ (mTOS, also quorum sensing type) and (ii) each node uses its both $x$ and $y$ (dTOS), as shown in 
Fig.~\ref{1}(b) and Fig.~\ref{2}(b). It has been found that only former way (mTOS) could lead to 
TOS effect since the latter way (dTOS) behaves same as SOS (found analytically as well as numerically). 
Moreover, since threshold increases with the increase in number of oscillators, $\gamma_{syn}$ also increases as 
the side effect of mTOS along with the maximization of $\gamma_{dsyn}$. This situation is more clear in terms of a two node network (Fig.~\ref{2}) 
wherein SOS is the case of two oscillators (Fig.~\ref{2}(a)) whereas mTOS depicts the scenario of four oscillators (Fig.~\ref{2}(b2)).

To demonstrate the TOS effect in terms of MSF  ($\lambda(\sigma\rightarrow\infty)<0$) and drive-response system ($\lambda^c<0$) for scalar coupling, 
we use a ring network of size $N$ (Fig.~\ref{1}) 
and a two node network (Fig.~\ref{2}), respectively. Furthermore, it should be noted that 
TOS could be applied to any arbitrary network topology since TOS alters only the node dynamics not the 
network structure, e.g. a ring network of TOS has same $R$ as for SOS, i.e. 
$R=1/ sin^2(\pi/N)$ where $\delta^{ring}_{min}=4sin^2(\pi/N)$  and $\delta^{ring}_{max}=4$ ($N$ is even~\cite{pecorapre}).

\subsubsection{MSF in case of TOS}
The equations of motion for the nodes of a ring network having TOS dynamics (Fig.~\ref{1}(b)), are:\newline
$\dot{x}^i=F(x^i,\mu)+\gamma_1\Gamma_{v}(y^i -x^i )$\newline
\hspace*{.6cm}$+\text{\boldmath$\gamma$}{\bf \Gamma(x^{i-1}-2x^{i}+x^{i+1})}$\newline
$\dot{y}^i=F(y^i,\mu')+\gamma_2\Gamma_{v}(x^i -y^i )$\hspace*{3.6cm}(4)\newline
\hspace*{.6cm}$+\text{\boldmath$\gamma$}{\bf \Gamma(y^{i-1}-2y^{i}+y^{i+1})}$\newline
~\newline
 $\dot{x}^i=F(x^i,\mu)+\gamma_1\Gamma_{v}(y^i -x^i )$\newline
$\dot{y}^i=F(y^i,\mu')+\gamma_2\Gamma_{v}(x^i -y^i )$\hspace*{3.6cm}(5)\newline
\hspace*{.6cm}$+\text{\boldmath$\gamma$}{\bf \Gamma(y^{i-1}-2y^{i}+y^{i+1})}$\newline
Corresponding to Fig.~\ref{1}(b1) and (b2), Eq. 4 and Eq. 5 respectively, represent the scenarios of 
N identical diffusively coupled dTOS and mTOS. The bold terms of Eq. 4-5 show the nearest 
neighborhood interactions form of $\gamma\sum\limits_{j=1}^{N}G_{ij}H(x^j)$ (Eq. 1) 
wherein $\Gamma$ represents scalar coupling scenario (same as Eq. 3).

To understand the different behavior of dTOS and mTOS, consider 
the block diagonalized linear variational equations of Eq. 4-5 by using technique of 
spatial Fourier modes (given in ref.~\cite{heagypre}):\newline 
$\dot{\eta}_x^k=(DF(s,\mu)-\gamma_1\Gamma_{v}{\bf-4}\text{\boldmath$\gamma$}{\bf sin^2(\pi k/N)\Gamma})\eta_x^k\newline\hspace*{.6cm}+\gamma_1\Gamma_{v}\eta_y^k$\newline
$\dot{\eta}_y^k=(DF(y_s,\mu')-\gamma_2\Gamma_{v}{\bf-4}\text{\boldmath$\gamma$}{\bf sin^2(\pi k/N)\Gamma})\eta_y^k
\newline\hspace*{.6cm}+\gamma_2\Gamma_{v}\eta_x^k$\hspace*{5.9cm}(6)\newline
~\newline 
$\dot{\eta}_x^k=(DF(s,\mu)-\gamma_1\Gamma_{v})\eta_x^k+\gamma_1\Gamma_{v}\eta_y^k$\newline
$\dot{\eta}_y^k=(DF(y_s,\mu')-\gamma_2\Gamma_{v}{\bf-4}\text{\boldmath$\gamma$}{\bf sin^2(\pi k/N)\Gamma})\eta_y^k
\newline\hspace*{.6cm}+\gamma_2\Gamma_{v}\eta_x^k$\hspace*{5.9cm}(7)\newline  
Where $\eta_x^k=(1/N)\sum\limits_{j=0}^{N-1}\xi_x^je^{2\pi ijk/N}$, $\eta_y^k=(1/N)\sum\limits_{j=0}^{N-1}\xi_y^je^{2\pi ijk/N}$ and 
$\xi_x^i=x^{i}-s$, $\xi_y^i=y^{i}-y_s$. Moreover, $DF(s,\mu)$ and $DF(y_s,\mu')$ are the Jacobian matrices 
evaluated at the synchronization manifold: $x=s$, $y=y_s$; $\dot{s}=F(s,\mu)+\gamma_1\Gamma_{v}(y_s-s)$, 
$\dot{y_s}=F(y_s,\mu')+\gamma_2\Gamma_{v}(s-y_s)$. The bold terms of Eq. 6-7 depict the form of  $\gamma \delta_kDH(s)$ (Eq. 2) in case of 
ring network topology. Eq. 6 and 7 are corresponding to Eq. 4 (dTOS) and 5 (mTOS), respectively. 
By using $\eta'^k=\gamma_2\eta_x^k+\gamma_1\eta_y^k$, 
Eq. 6 and Eq. 7 become:\newline
$\dot{\eta}'^k=DF(s,\mu)\gamma_2\eta_x^k+DF(y_s,\mu')\gamma_1\eta_y^k\newline\hspace*{.6cm}{\bf-4}\text{\boldmath$\gamma$}{\bf sin^2(\pi k/N)\Gamma}\eta'^k$\hspace*{4.3cm}(8)\newline
~\newline
$\dot{\eta}'^k=DF(s,\mu)\gamma_2\eta_x^k+DF(y_s,\mu')\gamma_1\eta_y^k\newline\hspace*{.6cm}{\bf-4}\text{\boldmath$\gamma$}{\bf sin^2(\pi k/N)\Gamma}\eta'^k+\gamma_2{\bf 4}\text{\boldmath$\gamma$}{\bf sin^2(\pi k/N)\Gamma}\eta_x^k$\newline
$\dot{\eta}_x^k=(DF(s,\mu)-(\gamma_1+\gamma_2)\Gamma_{v})\eta_x^k+\Gamma_{v}\eta'^k$\hspace*{2.07cm}(9)\newline
To simplify, we assume $DF(s,\mu)\approx DF(y_s,\mu')=D$ since TOS due to vector coupling ($\Gamma_{v}$) and 
strong values of intra-coupling constants ($\gamma_1$,$\gamma_2$), 
could behave as a system whose divergence rates from the synchronization manifold (complete/generalized) for both the oscillators 
could be equal even when $\delta\mu\neq0$ ($\delta\mu=\mu'-\mu$). This assumption is valid for every oscillator model for a 
given range of $\delta\mu$ (similar type of assumption had also been used previously~\cite{pecorapar}). Therefore using this assumption, 
Eq. 8 (dTOS) becomes:\newline
$\dot{\eta}'^k=(D{\bf-4}\text{\boldmath$\gamma$}{\bf sin^2(\pi k/N)\Gamma})\eta'^k$\hspace*{3.3cm}(10)\newline 
and Eq. 9 (mTOS) becomes:\newline
$\dot{\eta}'^k=(D{\bf-4}\text{\boldmath$\gamma$}{\bf sin^2(\pi k/N)\Gamma})\eta'^k\newline
\hspace*{.6cm}+\gamma_2{\bf 4}\text{\boldmath$\gamma$}{\bf sin^2(\pi k/N)\Gamma}\eta_x^k$\hspace*{3.6cm}(11a)\newline
$\dot{\eta}_x^k=(D-(\gamma_1+\gamma_2)\Gamma_{v})\eta_x^k+\Gamma_{v}\eta'^k$\hspace*{3cm}(11b)\newline
The form of Eq. 10 evidently shows that dTOS behaves same as SOS (Eq. 2). On the contrary, Eq. 11 depicts that in case of 
mTOS, the perturbation $\eta$ also depends upon its $x$-component ($\eta_x$) in addition to coupling strength ($\gamma$) and eigenmodes ($k$). Moreover, 
since the evolution of $\eta_x$ does not depend on $\gamma$ and $k$ (Eq. 11b), it could act as a stability factor. However, 
this analysis (without numerics) does not provide any hint that $\gamma$-independent Eq. 11b could lead to the maximization of $\gamma_{dsyn}$. 
Thus, we need to further investigate by using drive-response framework. 

Furthermore, it should be noted that since each $k$ is twice degenerate in case of ring network,
MSF ($\lambda(\gamma\delta_{max})$) is obtained by solving the perturbation equations (Eq. 10-11) for 
$k=N/2$ (maximum eigenvalue), i.e. $\lambda(4\gamma)=\lambda(\gamma\delta_{max})$. 

 \subsubsection{Drive-Response in case of TOS}
 The equations of motion for two unidirectionally coupled identical nodes having TOS dynamics are:
$\dot{x}^1=F(x^1,\mu)+\gamma_1\Gamma_{v}(y^1 -x^1 )$, 
$\dot{y}^1=F(y^1,\mu')+\gamma_2\Gamma_{v}(x^1 -y^1 )$\newline
~\newline
$\dot{x}^2=F(x^2,\mu)+\gamma_1\Gamma_{v}(y^2 -x^2 )$\newline
\hspace*{.6cm}$+\text{\boldmath$\gamma$}{\bf \Gamma(x^{1}-x^{2}})$\newline
$\dot{y}^2=F(y^2,\mu')+\gamma_2\Gamma_{v}(x^2 -y^2 )$\newline
\hspace*{.6cm}$+\text{\boldmath$\gamma$}{\bf \Gamma(y^{1}-y^{2}})$\hspace*{5.1cm}(12)\newline
~\newline
$\dot{x}^2=F(x^2,\mu)+\gamma_1\Gamma_{v}(y^2 -x^2 )$\newline
$\dot{y}^2=F(y^2,\mu')+\gamma_2\Gamma_{v}(x^2 -y^2 )$\newline
\hspace*{.6cm}$+\text{\boldmath$\gamma$}{\bf \Gamma(y^{1}-y^{2}})$\hspace*{5.1cm}(13)\newline
Similar to Eq. 3, the node $x^1-y^1$ is drive whereas the node $x^2-y^2$ is response. Eq. 12 and Eq. 13 
depict the scenarios of dTOS and mTOS, respectively. Now we choose $x_1$, $y_1$ coordinates for dTOS whereas 
$y_1$ coordinate in case of mTOS for the interactions and assume that the coupling term vanishes at $\gamma\rightarrow\infty$. 
Hence, Eq. 12 and 13 become Eq. 14-15 and Eq. 16-17, respectively (same as discussed for Eq. 3):\newline
$\dot{x}^2=F(x^2,x^1_1,\mu)+\gamma_1\Gamma_{v}(y^2 -x^2 )$\hspace*{2.9cm}(14)\newline
$\dot{y}^2=F(y^2,y^1_1,\mu')+\gamma_2\Gamma_{v}(x^2 -y^2 )$\hspace*{2.9cm}(15)\newline
~\newline
$\dot{x}^2=F(x^2,y^1_1,\mu)+\gamma_1\Gamma_{v}(y^2 -x^2 )$\hspace*{2.9cm}(16)\newline
$\dot{y}^2=F(y^2,y^1_1,\mu')+\gamma_2\Gamma_{v}(x^2 -y^2 )$\hspace*{2.9cm}(17)\newline
Eq. 14-15 depict that for dTOS, $x^2$ and $y^2$ depend only on their respective missing components as 
$x^1_1$ only drives $x^2$ and $y^1_1$ only drives $y^2$ which is similar to Eq. 3. On the other hand, 
Eq. 16 shows that in case of mTOS both $x^2$ and $y^2$ are driven by same $y^1_1$. Now we need to  
analyze $y^1_1$ driving effect on $x^2$ (which is missing in dTOS). For this we decompose 
only Eq. 16 (because Eq. 17 is same as Eq. 15):\newline
$\dot{x}_1^2=f_1(x^2,\mu)+\gamma_1(y^1_1 -x_1^2 )$\hspace*{3.9cm}(18)\newline
$\dot{x}_i^2=f_i(x^2,\mu)+\gamma_1(y_i^2 -x_i^2 )$\hspace*{3.9cm}(19)\newline
Where $i=2,..,m$. The relevance of  $y^1_1$-drive would clearly emerge when one takes partial derivative of Eq. 18 
w.r.t. $x_1^2 $ which results into the addition of $-\gamma_1$ in the 
first element of sub-Jacobian whose counter balance term, i.e. $+\gamma_1$,  
is missing in the non diagonal element (i.e. $0$) because Eq. 17 does not have $\dot{y}_1^2$. Thus, this uncompensated $-\gamma_1$ 
becomes the source of TOS effect, i.e. $\lambda^c<0$ ($\lambda(\sigma\rightarrow\infty)<0$). This is in contrast to 
the sub-Jacobian for all other cases (Eq. 14-15, 17 and 19) wherein the negative term ($-\gamma_1/\gamma_2$) of the diagonal elements 
gets balanced by the positive terms ($\gamma_1/\gamma_2$) of non-diagonal elements. Therefore, we can say that Eq. 18 
demonstrates the role of $\gamma$-independent Eq. 11b towards the maximization of $\gamma_{dsyn}$. In addition, 
Eq. 18 also explains the need of condition 2 (Sec. I), i.e. $x_1$ ($y_1$) should stabilize the unstable fixed point of $x$ ($y$) 
since the TOS effect emerges only due to the additional linear dissipative term ($-\gamma_1x_1$). Moreover, this also 
shows the relevance of $\gamma_1$, i.e. to induce the TOS effect 
$\gamma_{1_{01}}\leq\gamma_1\leq\gamma_{1_{02}}$ where $\gamma_{1_{01}}$ and $\gamma_{1_{02}}$ 
depend on the choice of oscillator model. In the present work,  the lower bound $\gamma_{1_{01}}$ has been found 
by exploiting the quorum sensing conditions~\cite{qs} (the upper bound $\gamma_{1_{02}}$ has been 
located just by scanning the intra-coupling parameter space).

Furthermore, it should also be noted that in case of scalar coupling (like $\Gamma$=diag(1,0..,0)$_{m\times m}$), 
the sub-Jacobian matrices sizes for Eq. 3 (SOS), Eq. 12 (dTOS) and Eq. 13 (mTOS) are:  $(m-1)\times (m-1)$, 
$(2m-2)\times (2m-2)$ and  $(2m-1)\times (2m-1)$, respectively.

\subsection{Oscillator Models}
\subsubsection{Chaotic}
R{\"o}ssler oscillator~\cite{rossler}: $F(x,\mu)=[-x_2-x_3;x_1+0.15x_2;0.15+x_3(x_1-d) ]'$,
$F(y,\mu')=[-y_2-y_3;y_1+\mu_Ry_2;\mu_R+y_3(y_1-d) ]'$ where $d$ and $\mu_R$ are the bifurcation parameters. 

Chua oscillator~\cite{chua}: $F(x,\mu)=[8.5(x_2-x_1-g(x_1));x_1-x_2+x_3;-14.97x_2 ]'$ (chaotic behavior), 
$F(y,\mu')=[\mu_C(y_2-y_1-g(y_1));y_1-y_2+y_3;-14.97y_2 ]'$ where 
$\mu_C$ is the bifurcation parameter and $g(x_1/y_1)=mx_1/y_1+0.5(m_0-m)(abs(x_1/y_1+1)-abs(x_1/y_1-1))$ 
($m$=-0.68;$m_0$=-1.31). 

Lorenz oscillator~\cite{lorenz}: 
$F(x,\mu)=[10(x_2-x_1);-x_1x_3+28x_1-x_2;x_1x_2-2x_3 ]'$ (chaotic behavior), 
$F(y,\mu')=[10(y_2-y_1);-y_1y_3+28y_1-y_2;y_1y_2-\mu_Ly_3]'$ where 
$\mu_L$ is the bifurcation parameter. 
\subsubsection{Hyperchaotic}
The employed four dimensional electronic system~\cite{pecorahyper} (piecewise linear form of hyperchaotic R{\"o}ssler~\cite{rosslerhyper}): 
$F(x,\mu)=[-0.05x_1-0.5x_2-0.62x_3;x_1+0.15 x_2+0.40x_4;-2x_3+f(x_1);-1.5x_3+0.18x_4+h(x_4) ]'$,
$F(y,\mu')=[-0.05y_1-0.5y_2-0.62y_3;y_1+0.15 y_2+0.40y_4;-2y_3+f(y_1);-1.5y_3+0.18y_4+h(y_4) ]'$ 
where $f(x_1/y_1)=10(x_1/y_1-0.68)\Theta(x_1/y_1-0.68)$ and 
$h(x_4/y_4)=-0.41(x_4/y_4-3.8)\Theta(x_4/y_4-3.8)$. Here, $\Theta(\cdot)$ is the Heaviside step function.

\begin{figure}
\includegraphics[height=4.5cm,width=8cm,angle=0]{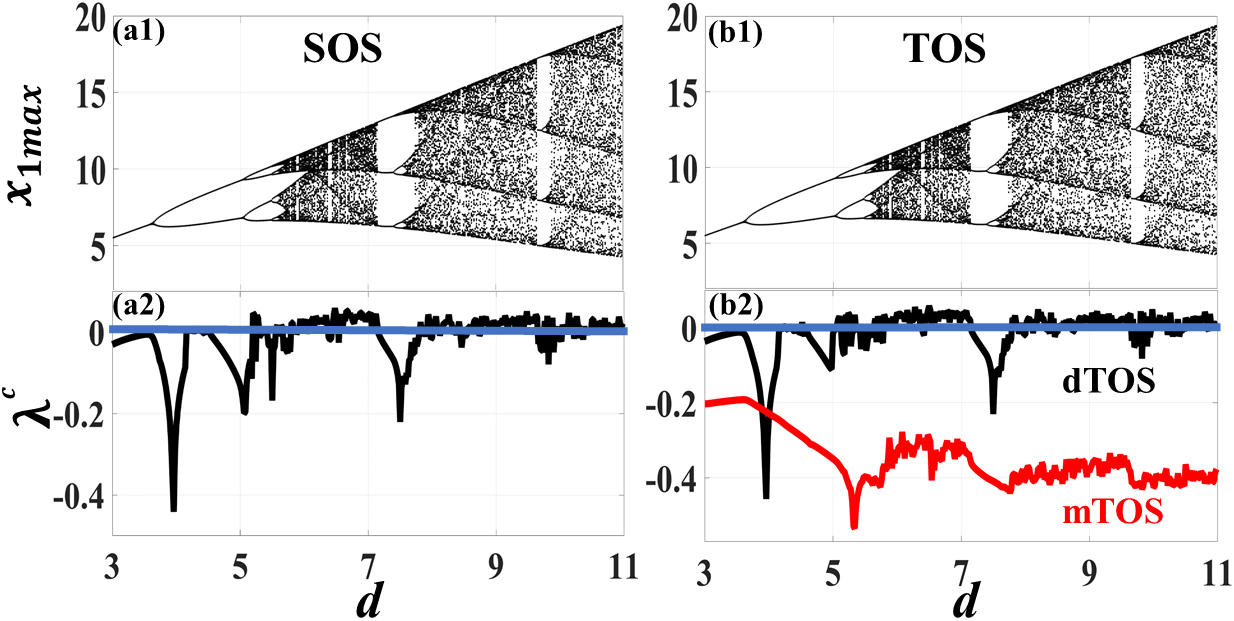}
\caption{(Color online) For $x_2$-driving R{\"o}ssler, plots depict the stability of two unidirectionally coupled identical nodes (Fig.~\ref{2}) 
using $\lambda^c$ (a2-b2) at various dynamical regimes of R{\"o}ssler (a1-b1) generated by its bifurcation parameter $d$. These 
plots demonstrate that the robustness emerges only for mTOS configuration, i.e. $\lambda^c<0$ for all $d$ values. Here, the 
used TOS parameters are: $\gamma_1$=$\gamma_2$=$2$ and 
the model parameter of $y$ is: $\mu_R$=$0.15$ (i.e. $y$=$x$, identical TOS scenario). Solid blue line depicts 
zero base line, i.e. at which $\lambda^c=0$.} 
\label{3}
\end{figure}

\begin{figure}
\includegraphics[height=3.36cm,width=8cm,angle=0]{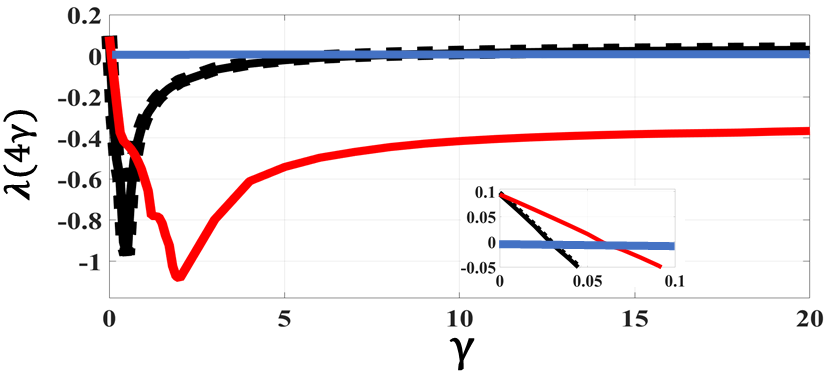}
\caption{(Color online) Corresponding to $d=6.8$ of Fig.~\ref{3}, plot depicts the stability of a ring network of $N$ identical nodes 
(Fig.~\ref{1}) using $\lambda(4\gamma)$ (MSF) with the increase in coupling strength $\gamma$ for 
SOS (dotted black), dTOS (solid black) and mTOS (solid red) scenarios. Figure shows that 
mTOS maximizes $\gamma_{dsyn}$ whereas dTOS behaves qualitatively same as SOS, i.e. small $\gamma_{dsyn}$ (=6.01).  
Inset plot (magnified version) represents the behavior of 
MSF at small values of $\gamma$ which shows the increment of $\gamma_{syn}$ (i.e. from 0.03 to 0.06) in case of mTOS. 
Solid blue line depicts 
zero base line, i.e. at which $\lambda(4\gamma)=0$.} 
\label{4}
\end{figure}

\begin{figure}
\includegraphics[height=3.36cm,width=8cm,angle=0]{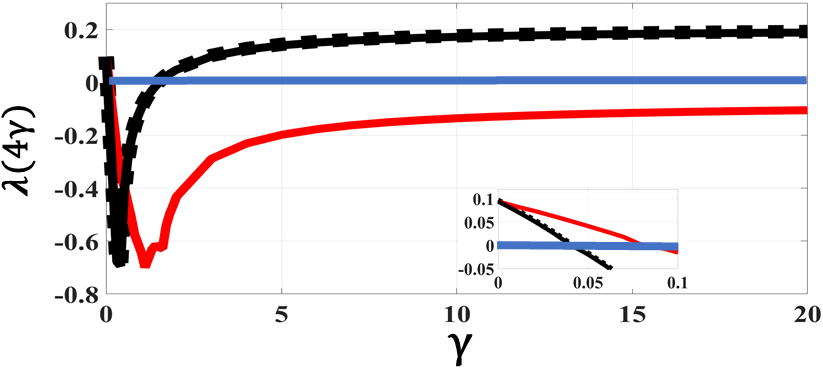}
\caption{(Color online) For $x_1$-coupled chaotic R{\"o}ssler, plot depicts the stability of a ring network of $N$ identical nodes 
(Fig.~\ref{1}) using $\lambda(4\gamma)$ (MSF) with the increase in coupling strength $\gamma$ for 
SOS (dotted black), dTOS (solid black) and mTOS (solid red) scenarios. Figure shows that 
mTOS maximizes $\gamma_{dsyn}$ whereas dTOS behaves qualitatively same as SOS, i.e. small $\gamma_{dsyn}$ (=1.5).  
Inset plot (magnified version) represents the behavior of 
MSF at small values of $\gamma$ which shows the increment of $\gamma_{syn}$ (i.e. from 0.04 to 0.09) in case of mTOS. The parameters 
used in this scenario are same as Fig.~\ref{3}-\ref{4} (i.e. identical TOS). Solid blue line depicts 
zero base line, i.e. at which $\lambda(4\gamma)=0$.} 
\label{5}
\end{figure}

\begin{figure}
\includegraphics[height=5cm,width=8cm,angle=0]{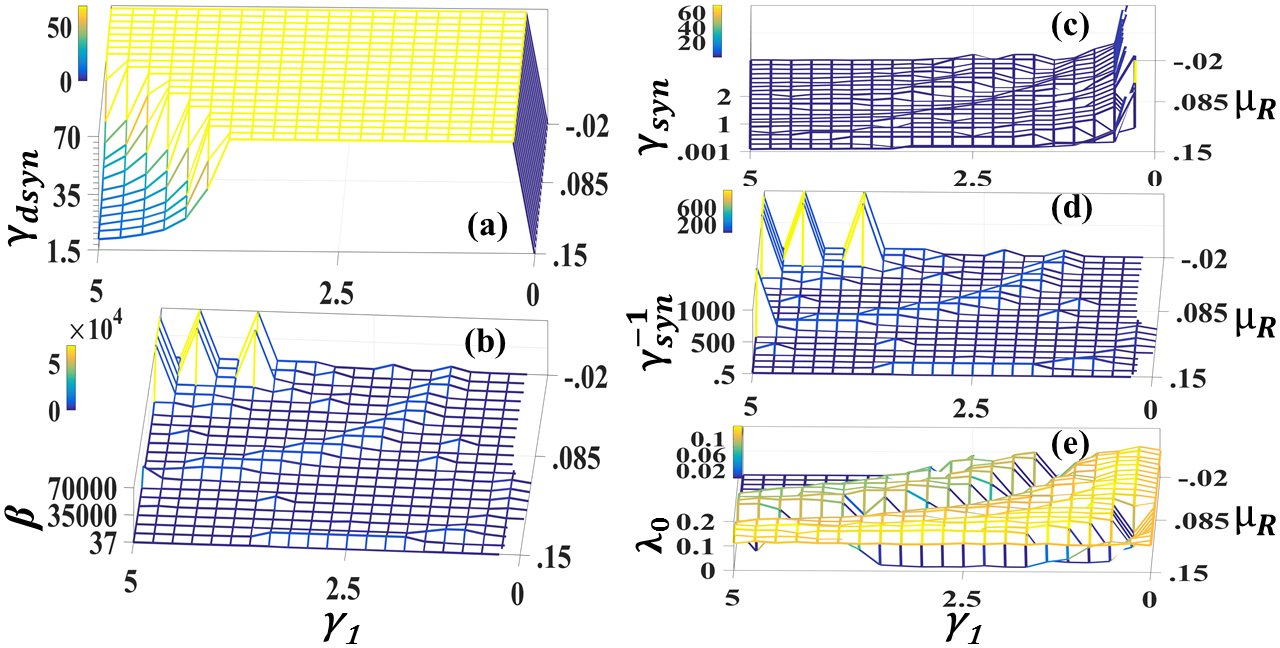}
\caption{(Color online) Plot depicts the robustness and the side effect of mTOS 
in case of Fig.~\ref{5} at fixed $\gamma_2$=5 (no-feedback), i.e. 
the alterations of $\gamma_{dsyn}$ (a), $\beta$ (b), $\gamma_{syn}$ (c), $\gamma^{-1}_{syn}$ 
(d) and $\lambda_{0}$~\cite{tos} (e) against the variation of 
TOS parameter $\gamma_1$ and the model parameter $\mu_R$. In subplot (a), the yellow
region implies mTOS effect ($\lambda(4\gamma\rightarrow\infty)<0$), 
blue region shows weak-mTOS effect ($\gamma_{dsyn}>$1.5, $\lambda(4\gamma\rightarrow\infty)>0$) and 
no-mesh region represents no-mTOS effect ($\gamma_{dsyn}\leq$1.5, i.e. at $\gamma_1<\gamma_{1_{01}}$). In subplot (b), yellow region 
means strong-enhancement ($\beta>>$37) and blue region represents $\beta$ enhancement ($\beta>$37) 
where no-mesh region depicts no-enhancement ($\beta\leq$37). Subplot (c) depicts the incrementation of 
$\gamma_{syn}$ via mTOS (no-mesh for large $\gamma_{syn}$ 
as $\gamma_1\rightarrow0$, $\gamma_{syn}\rightarrow\infty$).  
Since $\beta=\gamma_{dsyn}\gamma_{syn}^{-1}$, subplot (d) 
demonstrates that the enhancement as well as no-enhancement regions of subplot (b) 
depend on the variation of $\gamma_{syn}$. Subplots (e) and (a) show that the state of 
synchronization manifold $\lambda_{0}$ does not effect $\gamma_{dsyn}$.} 
\label{6}
\end{figure}

\begin{figure}
\includegraphics[height=5cm,width=8cm,angle=0]{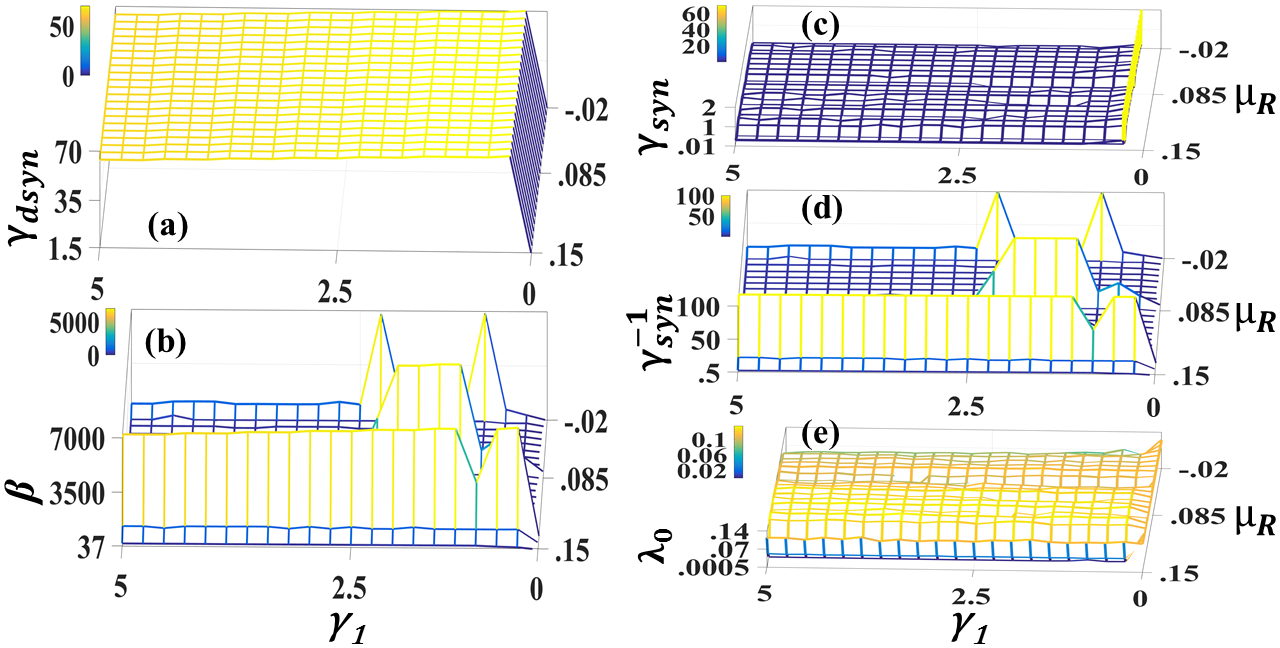}
\caption{(Color online) Plot depicts the robustness and the side effect of mTOS 
in case of Fig.~\ref{5} at varying $\gamma_2$=5$\gamma_1$ (feedback), i.e. 
the alterations of $\gamma_{dsyn}$ (a), $\beta$ (b), $\gamma_{syn}$ (c), $\gamma^{-1}_{syn}$ 
(d) and $\lambda_{0}$~\cite{tos} (e) against the variation of 
TOS parameter $\gamma_1$ and the model parameter $\mu_R$. In subplot (a), the yellow
region implies mTOS effect ($\lambda(4\gamma\rightarrow\infty)<0$), 
blue region shows weak-mTOS effect ($\gamma_{dsyn}>$1.5, $\lambda(4\gamma\rightarrow\infty)>0$) and 
no-mesh region represents no-mTOS effect ($\gamma_{dsyn}\leq$1.5, i.e. at $\gamma_1<\gamma_{1_{01}}$). In subplot (b), yellow region 
means strong-enhancement ($\beta>>$37) and blue region represents $\beta$ enhancement ($\beta>$37) 
where no-mesh region depicts no-enhancement ($\beta\leq$37). Subplot (c) depicts the incrementation of 
$\gamma_{syn}$ via mTOS (no-mesh for large $\gamma_{syn}$ 
as $\gamma_1\rightarrow0$, $\gamma_{syn}\rightarrow\infty$).  
Since $\beta=\gamma_{dsyn}\gamma_{syn}^{-1}$, subplot (d) 
demonstrates that the enhancement as well as no-enhancement regions of subplot (b) 
depend on the variation of $\gamma_{syn}$. Subplots (e) and (a) show that the state of 
synchronization manifold $\lambda_{0}$ does not effect $\gamma_{dsyn}$.} 
\label{7}
\end{figure}

\begin{figure}
\includegraphics[height=3.36cm,width=8cm,angle=0]{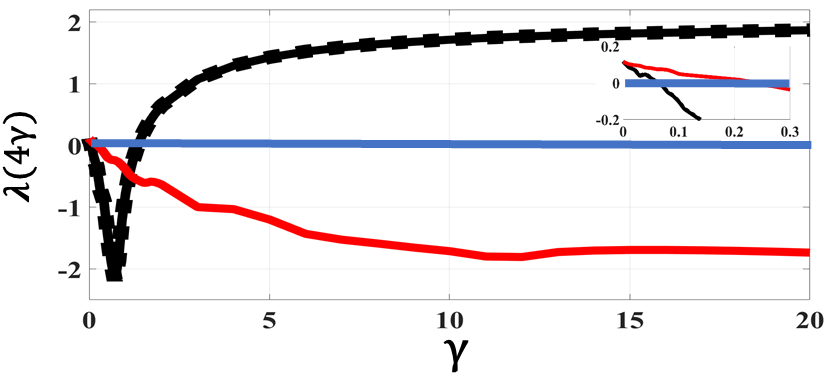}
\caption{(Color online) For $x_3$-coupled chaotic Chua, plot depicts the stability of a ring network of $N$ identical nodes 
(Fig.~\ref{1}) using $\lambda(4\gamma)$ (MSF) with the increase in coupling strength $\gamma$ for 
SOS (dotted black), dTOS (solid black) and mTOS (solid red) scenarios. Figure shows that 
mTOS maximizes $\gamma_{dsyn}$ whereas dTOS behaves qualitatively same as SOS, i.e. small $\gamma_{dsyn}$ (=1.30).  
Inset plot (magnified version) represents the behavior of 
MSF at small values of $\gamma$ which shows the increment of $\gamma_{syn}$ (i.e. from 0.09 to 0.31) in case of mTOS. Here, the 
used TOS parameters are: $\gamma_1$=2, $\gamma_2$=5 and 
the model parameter of $y$ is: $\mu_C$=$8.5$ (i.e. $y$=$x$, identical TOS scenario). Solid blue line depicts 
zero base line, i.e. at which $\lambda(4\gamma)=0$.} 
\label{8}
\end{figure}

\begin{figure}
\includegraphics[height=5cm,width=8cm,angle=0]{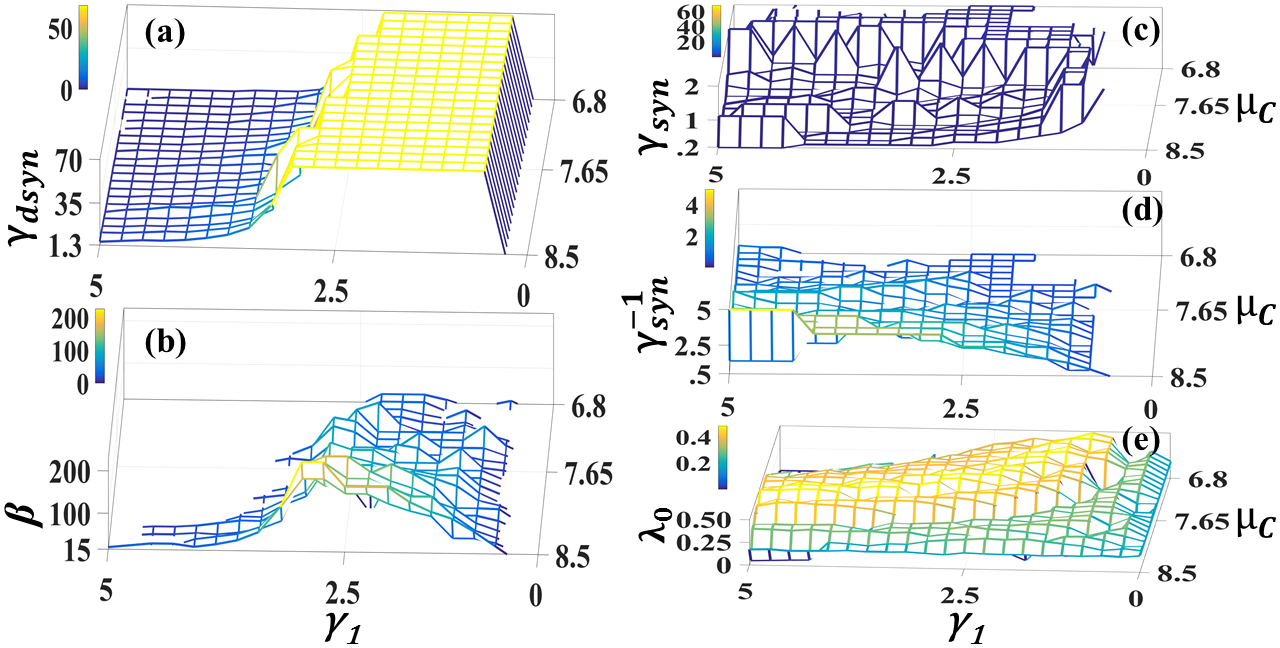}
\caption{(Color online) Plot depicts the robustness and the side effect of mTOS 
in case of Fig.~\ref{8} at fixed $\gamma_2$=5 (no-feedback), i.e. 
the alterations of $\gamma_{dsyn}$ (a), $\beta$ (b), $\gamma_{syn}$ (c), $\gamma^{-1}_{syn}$ 
(d) and $\lambda_{0}$~\cite{tos} (e) against the variation of 
TOS parameter $\gamma_1$ and the model parameter $\mu_C$. In subplot (a), the yellow
region implies mTOS effect ($\lambda(4\gamma\rightarrow\infty)<0$), 
blue region shows weak-mTOS effect ($\gamma_{dsyn}>$1.3, $\lambda(4\gamma\rightarrow\infty)>0$) and 
no-mesh region represent no-mTOS effect ($\gamma_{dsyn}\leq1.3$, i.e. at $\gamma_1<\gamma_{1_{01}}$). In subplot (b), yellow and 
blue regions represent $\beta$ enhancement ($\beta>$15) where no-mesh region 
depicts no-enhancement ($\beta\leq$15). Subplot (c) depicts the incrementation of 
$\gamma_{syn}$ via mTOS (no-mesh for large $\gamma_{syn}$ as $\gamma_1\rightarrow0$, $\gamma_{syn}\rightarrow\infty$).  
Since $\beta=\gamma_{dsyn}\gamma_{syn}^{-1}$, subplot (d) 
demonstrates that the enhancement as well as no-enhancement regions of subplot (b) 
depend on the variation of $\gamma_{syn}$. Subplots (e) and (a) show that the state of 
synchronization manifold $\lambda_{0}$ does not effect $\gamma_{dsyn}$.} 
\label{9}
\end{figure}

\begin{figure}
\includegraphics[height=5cm,width=8cm,angle=0]{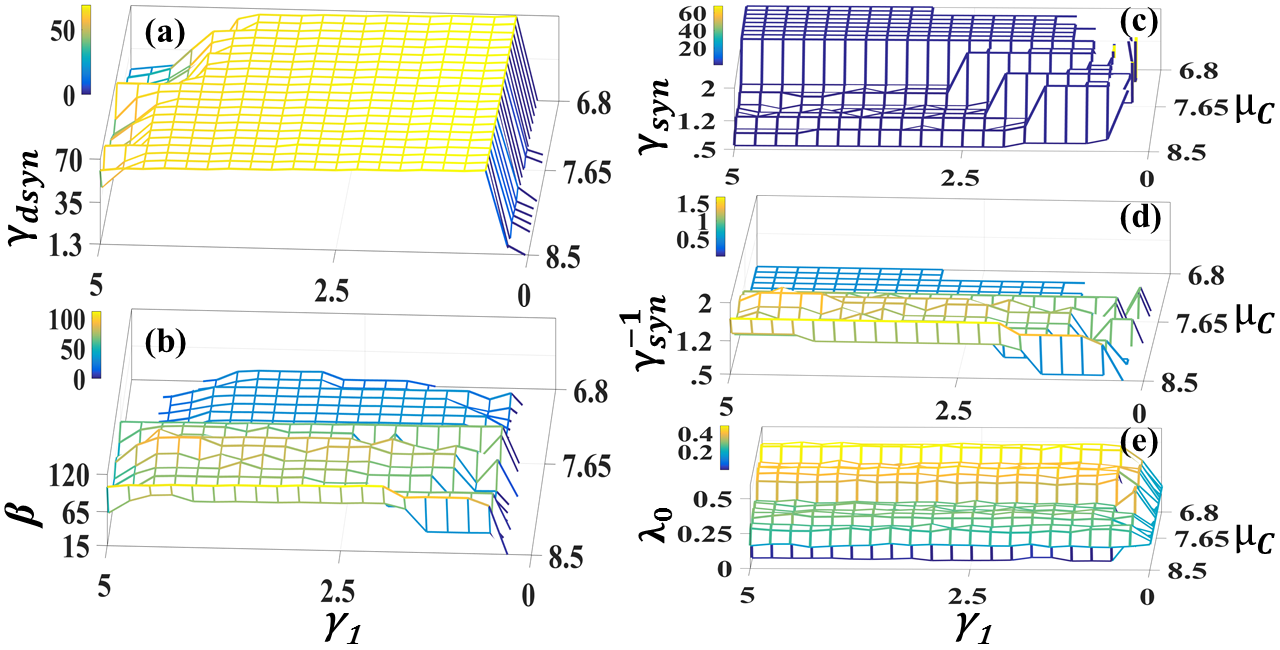}
\caption{(Color online) Plot depicts the robustness and the side effect of mTOS 
in case of Fig.~\ref{8} at varying $\gamma_2$=5$\gamma_1$ (feedback), i.e. 
the alterations of $\gamma_{dsyn}$ (a), $\beta$ (b), $\gamma_{syn}$ (c), $\gamma^{-1}_{syn}$ 
(d) and $\lambda_{0}$~\cite{tos} (e) against the variation of 
TOS parameter $\gamma_1$ and the model parameter $\mu_C$. 
In subplot (a), the yellow
region implies mTOS effect ($\lambda(4\gamma\rightarrow\infty)<0$), 
blue region shows weak-mTOS effect ($\gamma_{dsyn}>$1.3, $\lambda(4\gamma\rightarrow\infty)>0$) and 
no-mesh region represent no-mTOS effect ($\gamma_{dsyn}\leq1.3$, i.e. at $\gamma_1<\gamma_{1_{01}}$). In subplot (b), yellow and 
blue regions represent $\beta$ enhancement ($\beta>$15) where no-mesh region 
depicts no-enhancement ($\beta\leq$15). Subplot (c) depicts the incrementation of 
$\gamma_{syn}$ via mTOS (no-mesh for large $\gamma_{syn}$ as $\gamma_1\rightarrow0$, $\gamma_{syn}\rightarrow\infty$).  
Since $\beta=\gamma_{dsyn}\gamma_{syn}^{-1}$, subplot (d) 
demonstrates that the enhancement as well as no-enhancement regions of subplot (b) 
depend on the variation of $\gamma_{syn}$. Subplots (e) and (a) show that the state of 
synchronization manifold $\lambda_{0}$ does not effect $\gamma_{dsyn}$.} 
\label{10}
\end{figure}

\begin{figure}
\includegraphics[height=3.36cm,width=8cm,angle=0]{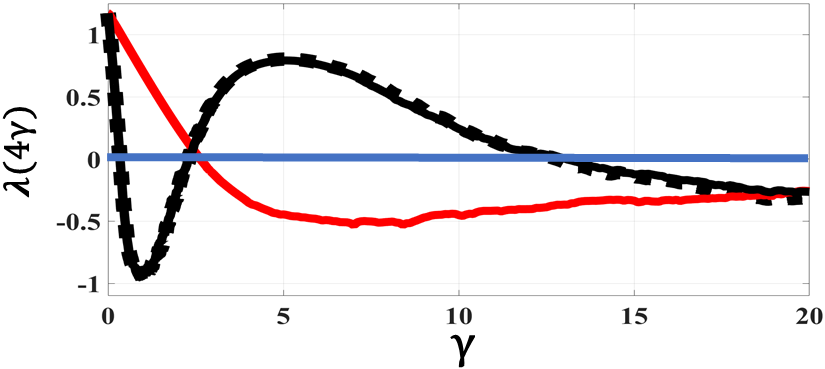}
\caption{(Color online) For $x_3$-coupled chaotic Lorenz, plot depicts the stability of a ring network of $N$ identical nodes 
(Fig.~\ref{1}) using $\lambda(4\gamma)$ (MSF) with the increase in coupling strength $\gamma$ for 
SOS (dotted black), dTOS (solid black) and mTOS (solid red) scenarios. Figure shows the stabilization via mTOS, i.e. 
the two values of $\gamma_{syn}$ (0.35 and 12.81) in the cases of SOS and dTOS, become a single value (=2.63) and 
first $\gamma_{dsyn}$ (2.4) is maximized. Here, the used TOS parameters are: $\gamma_1$=4, $\gamma_2$=20 and 
the model parameter of $y$ is: $\mu_L$=$2$ (i.e. $y$=$x$, identical TOS scenario). Solid blue line depicts 
zero base line, i.e. at which $\lambda(4\gamma)=0$.} 
\label{11}
\end{figure}

\begin{figure}
\includegraphics[height=5cm,width=8cm,angle=0]{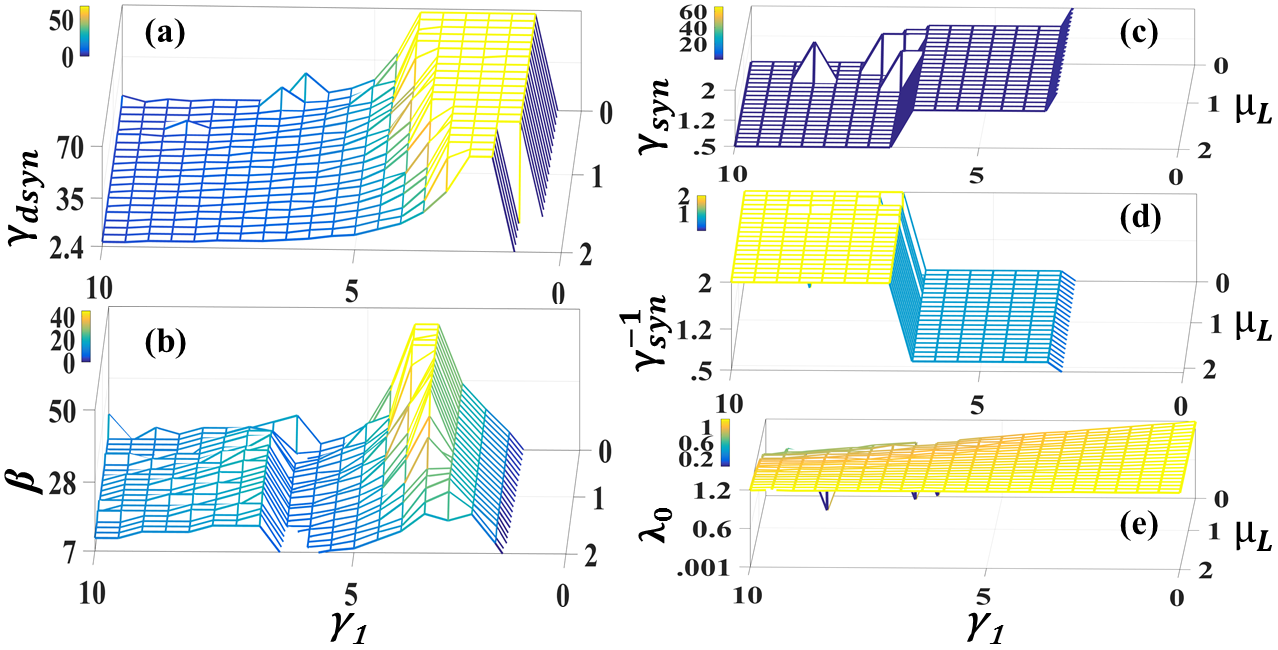}
\caption{(Color online) Plot depicts the robustness and the side effect of mTOS 
in case of Fig.~\ref{11} at fixed $\gamma_2$=10 (no-feedback), i.e. 
the alterations of $\gamma_{dsyn}$ (a), $\beta$ (b), $\gamma_{syn}$ (c), $\gamma^{-1}_{syn}$ 
(d) and $\lambda_{0}$~\cite{tos} (e) against the variation of 
TOS parameter $\gamma_1$ and the model parameter $\mu_L$. In subplot (a), the yellow
region implies mTOS effect ($\lambda(4\gamma\rightarrow\infty)<0$), 
blue region shows weak-mTOS effect ($\gamma_{dsyn}>$2.4, $\lambda(4\gamma\rightarrow\infty)>0$) and 
no-mesh region represent no-mTOS effect ($\gamma_{dsyn}\leq2.4$, i.e. at $\gamma_1<\gamma_{1_{01}}$). In subplot (b), yellow and 
blue regions represent $\beta$ enhancement ($\beta>$7) where no-mesh region 
depicts no-enhancement ($\beta\leq$7). Subplot (c) depicts the incrementation of 
$\gamma_{syn}$ via mTOS ((no-mesh for large $\gamma_{syn}$ as $\gamma_1\rightarrow0$, $\gamma_{syn}\rightarrow\infty$).  
Since $\beta=\gamma_{dsyn}\gamma_{syn}^{-1}$, subplot (d) 
demonstrates that the enhancement as well as no-enhancement regions of subplot (b) 
depend on the variation of $\gamma_{syn}$. Subplots (e) and (a) show that the state of 
synchronization manifold $\lambda_{0}$ does not effect $\gamma_{dsyn}$.} 
\label{12}
\end{figure}
 
\begin{figure}
\includegraphics[height=5cm,width=8cm,angle=0]{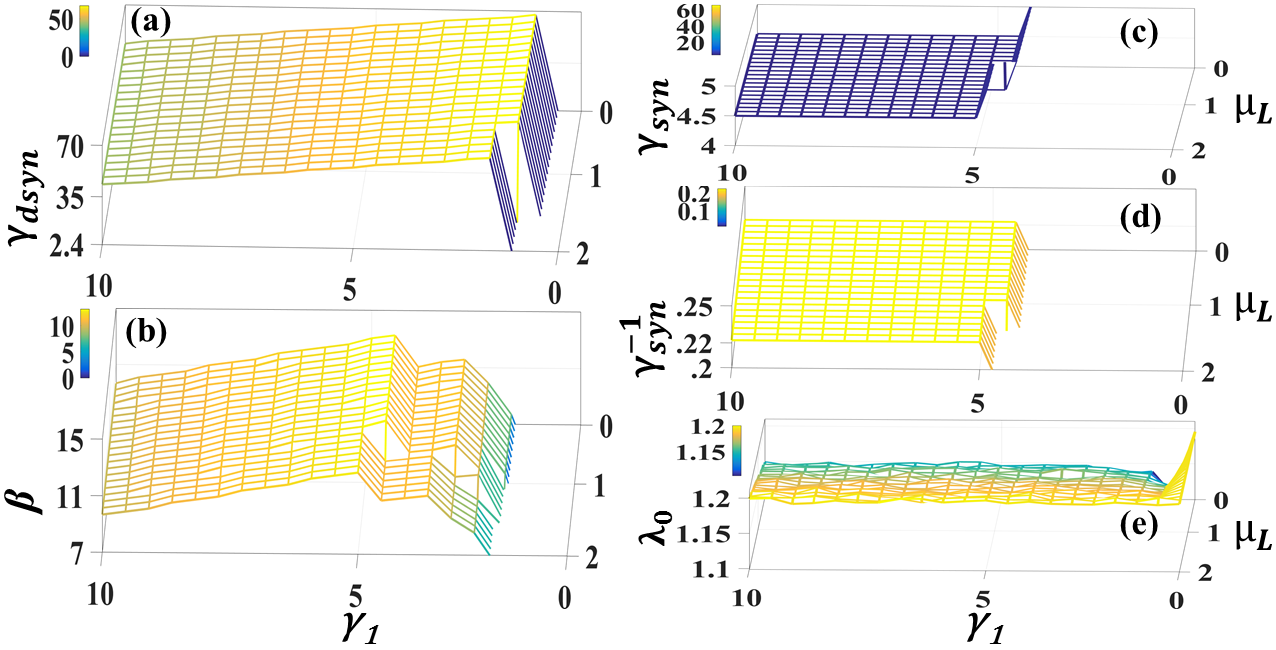}
\caption{Plot depicts the robustness and the side effect of mTOS 
in case of Fig.~\ref{11} at varying $\gamma_2$=10$\gamma_1$ (feedback), i.e. 
the alterations of $\gamma_{dsyn}$ (a), $\beta$ (b), $\gamma_{syn}$ (c), $\gamma^{-1}_{syn}$ 
(d) and $\lambda_{0}$~\cite{tos} (e) against the variation of 
TOS parameter $\gamma_1$ and the model parameter $\mu_L$. In subplot (a), the yellow
region implies mTOS effect ($\lambda(4\gamma\rightarrow\infty)<0$), 
blue region shows weak-mTOS effect ($\gamma_{dsyn}>$2.4, $\lambda(4\gamma\rightarrow\infty)>0$) and 
no-mesh region represent no-mTOS effect ($\gamma_{dsyn}\leq2.4$, i.e. at $\gamma_1<\gamma_{1_{01}}$). In subplot (b), yellow and 
blue regions represent $\beta$ enhancement ($\beta>$7) where no-mesh region 
depicts no-enhancement ($\beta\leq$7). Subplot (c) depicts the incrementation of 
$\gamma_{syn}$ via mTOS ((no-mesh for large $\gamma_{syn}$ as $\gamma_1\rightarrow0$, $\gamma_{syn}\rightarrow\infty$).  
Since $\beta=\gamma_{dsyn}\gamma_{syn}^{-1}$, subplot (d) 
demonstrates that the enhancement as well as no-enhancement regions of subplot (b) 
depend on the variation of $\gamma_{syn}$. Subplots (e) and (a) show that the state of 
synchronization manifold $\lambda_{0}$ does not effect $\gamma_{dsyn}$.} 
\label{13}
\end{figure}

\begin{figure}
\includegraphics[height=3.36cm,width=8cm,angle=0]{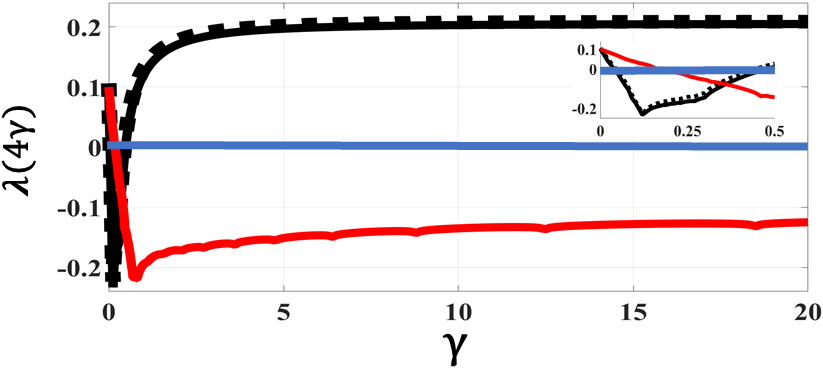}
\caption{(Color online) For $x_1$-coupled hyperchaotic circuit, plot depicts the stability of a ring network of $N$ identical nodes 
(Fig.~\ref{1}) using $\lambda(4\gamma)$ (MSF) with the increase in coupling strength $\gamma$ for 
SOS (dotted black), dTOS (solid black) and mTOS (solid red) scenarios. Figure shows that 
mTOS maximizes $\gamma_{dsyn}$ whereas dTOS behaves qualitatively same as SOS, i.e. small $\gamma_{dsyn}$ (=0.44).  
Inset plot (magnified version) represents the behavior of 
MSF at small values of $\gamma$ which shows the increment of $\gamma_{syn}$ (i.e. from 0.05 to 0.18) in case of mTOS. Here, the 
used TOS parameters are: $\gamma_1$=1, $\gamma_2$=3 ($y$=$x$, identical TOS scenario). Solid blue line depicts 
zero base line, i.e. at which $\lambda(4\gamma)=0$.} 
\label{14}
\end{figure}

\section{Results and Discussions}

In the present work, we employ 5 different scenarios: $x_1$ and $x_2$-coupled/driving chaotic R{\"o}ssler, $x_3$-coupled/driving chaotic Chua, 
$x_3$-coupled/driving chaotic Lorenz and $x_1$-coupled/driving hyperchaotic oscillator, 
to generalize the TOS effect, i.e. the stabilization of chaotic/hyperchaotic synchronization manifold ($\lambda^c<0$) or 
the maximization of $\gamma_{dsyn}$ ($\lambda(4\gamma\rightarrow\infty)<0$)
in case of scalar coupling. To reiterate, since these different scenarios implies different $DF$ and/or $DH$ (Eq. 2), each 
scenario has its own MSF. In addition, `coupled' term depicts explicit $\gamma$ dependence, i.e. the stability is shown by 
MSF using $\lambda(4\gamma)$ (ring network) whereas `driving' term shows the drive-response framework 
wherein the stabilization is given by maximum conditional Lyapunov exponent $\lambda^c$ of 
sub-Jacobian. Furthermore, the robustness of TOS effect has also been shown by varying the TOS parameters 
($\gamma_1$, $\gamma_2$) and the model parameters of R{\"o}ssler ($\mu_R$), Chua ($\mu_C$) and Lorenz ($\mu_L$).  
It has been found that the intra-coupling feedback situation (discussed in Sec. IIC) of the TOS parameters provides more robustness 
but lesser $\beta$ enhancement, in comparison to no feedback situation. 

It should be noted that the Lyapunov exponents are calculated by using Wolf's algorithm~\cite{wolf} and the model equations are integrated by using 
fourth order Runge-Kutta algorithm (integration step-size=0.01).

\subsection{R{\"o}ssler}
The emergent effect of TOS for R{\"o}ssler dynamics is given in Fig.~\ref{3}-\ref{7} wherein 
Fig.~\ref{3} depicts the stabilization of $x_2$-driving R{\"o}ssler using $\lambda^c$ against the variation of its 
bifurcation parameter $d$ for both SOS and TOS configurations. Fig.~\ref{3}(a) shows the scenario 
(which had also been previously reported in figure 10 of ref~\cite{pecorapra}) wherein
the subsystem ($x_2$-driven R{\"o}ssler) loses its stability close to the bifurcation points especially at the 
periodic windows in case of SOS. In contrast to Fig.~\ref{3}(a), Fig.~\ref{3}(b) evidently demonstrates that 
the mTOS configuration stabilizes the subsystem at all the domains whereas the dTOS configuration exhibits qualitatively similar 
behavior as of SOS (consistent with the theory). Moreover, Fig.~\ref{4} shows the 
MSF behavior for $x_2$-coupled chaotic R{\"o}ssler (for SOS and TOS) 
corresponding to $d=6.8$ (chaos) of Fig.~\ref{3} at which synchronization manifold 
is stable for mTOS ($\lambda^c<0$) and unstable for SOS ($\lambda^c>0$). 
Now it is evident from Fig.~\ref{4} that $\lambda(4\gamma\rightarrow\infty)<0$ only happens for mTOS configuration 
(where `$\infty$' means finitely large value). This also proves that at the large coupling strength MSF 
exhibits same information as the corresponding drive-response system does, i.e. 
$\lambda(4\gamma\rightarrow\infty)\approx \lambda^c$. Analogous to  Fig.~\ref{4},  
MSF behavior shown in Fig.~\ref{5} for $x_1$-coupled chaotic R{\"o}ssler also illustrates that only mTOS 
stabilizes the chaotic synchronization manifold or maximizes $\gamma_{dsyn}$. Moreover, the insets of Fig.~\ref{4} 
and~\ref{5} show the incrementation of $\gamma_{syn}$ as the side effect of mTOS configuration.

Furthermore, the mTOS effect shown in Fig.~\ref{5} is studied under parameter fluctuations, i.e. 
by varying TOS's critical parameter $\gamma_1$ (discussed in Sec. IIC2) and the model 
parameter $\mu_R$ (nonidentical TOS), for no-feedback (Fig.~\ref{6}) and feedback (Fig.~\ref{7}) scenarios. 
In contrast to Fig.~\ref{6}(a) wherein weak-mTOS effect observed at both small and 
large values of $\gamma_{1}$, Fig.~\ref{7}(a) evidently demonstrates the robustness of 
mTOS effect ($\lambda(4\gamma\rightarrow\infty)<0$) over the broader range 
of parameter space in case of feedback scenario. Moreover, it also suggests that for the induction of mTOS effect 
at large $\gamma_1$, $\gamma_{1}$ should be less than $\gamma_{2}$ (one can say that 
the upper bound of $\gamma_1$, $\gamma_{1_{02}}$, also depends on $\gamma_2$ along with the 
model parameters). However, the behavior of incrementation of $\gamma_{syn}$ increases with the increase in
$\gamma_{2}$ which reduces $\beta$ enhancement as depicted by 
Fig.~\ref{7}(b)-(d) in comparison to Fig.~\ref{6}(b)-(d) wherein $\gamma_{2}$ is fixed. Moreover, it should be noted that 
the extra strong-enhancement observed only in no-feedback scenario (Fig.~\ref{6}(b)) 
is the consequence of decrement of $\gamma_{syn}$ due to the stable periodic synchronization manifold, 
$\lambda_0=0$ (Fig.~\ref{6}(e)), which emerge at $\gamma_{1}\rightarrow\gamma_{2}$ for $\mu_R<0$ (R{\"o}ssler's steady state domain). 
But this behavior is missing in case of feedback scenario (Fig.~\ref{7}(b)) because $\gamma_2$ increases with the increase in 
$\gamma_1$ which results into chaotic dynamics, $\lambda_0>0$ (Fig.~\ref{7}(e)), even at $\mu_R<0$. 

Furthermore, it is worth noticing that analogous to Fig.~\ref{3}(b), 
the independence of $\gamma_{dsyn}$ maximization (Fig.~\ref{6}(a)-\ref{7}(a)) from the state of synchronization manifold (Fig.~\ref{6}(e)-\ref{7}(e)) 
clearly depict the potency of mTOS configuration. 

\subsection{Chua}
Analogous to Fig.~\ref{5}, Fig.~\ref{8} demonstrates the mTOS effect for $x_3$-coupled chaotic Chua and the robustness of this effect 
is given in Fig.~\ref{9}-\ref{10} (similar to Fig.~\ref{6}-\ref{7}). Here also, the feedback scenario  (Fig.~\ref{10})
provides more robustness but lesser $\beta$ enhancement than the case of no-feedback (Fig.~\ref{9}). Moreover, by comparing Fig.~\ref{9}(a) 
with Fig.~\ref{6}(a), one could realize the dependence of $\gamma_{1_{02}}$ on the oscillator model whereas Fig.~\ref{10}(a) depicts that 
this dependence gets reduced in the presence of feedback (similar to Fig.~\ref{7}(a)). But the major difference arises due to change in oscillator 
model from R{\"o}ssler to Chua, is the extent of $\beta$ enhancement as shown in Fig.~\ref{9}(b)-\ref{10}(b) which is lesser in contrast to Fig.~\ref{6}(b)-\ref{7}(b). 
This is because of more augmentation of $\gamma_{syn}$ in case of Chua (Fig.~\ref{9}(c)-(d) and~\ref{10}(c)-(d)) than the case of R{\"o}ssler 
(Fig.~\ref{6}(c)-(d) and~\ref{7}(c)-(d)). Instead of this discrepancy, Fig.~\ref{9}(a)-\ref{10}(a) in conjunction 
with Fig.~\ref{9}(e)-\ref{10}(e) clearly demonstrate the independence of $\gamma_{dsyn}$ maximization from the state of synchronization manifold which is 
similar to the case of R{\"o}ssler.

\subsection{Lorenz}
Similar to R{\"o}ssler and Chua, the case of $x_3$-coupled chaotic Lorenz given in Fig.~\ref{11}-\ref{13} also 
depict more robustness but lesser $\beta$ enhancement in case of feedback (Fig.~\ref{13}) than the case of no-feedback (Fig.~\ref{12}). In addition, 
consistent with R{\"o}ssler and Chua, Fig.~\ref{12}(a)-\ref{13}(a) in conjunction with Fig.~\ref{12}(e)-\ref{13}(e), evidently 
demonstrate the independence of $\gamma_{dsyn}$ maximization from the state of synchronization manifold. 
Again the difference emerge in the extent of $\beta$ enhancement (Fig.~\ref{12}(b)-\ref{13}(b)) which is much smaller in contrast to 
R{\"o}ssler because $\gamma_{syn}$ augments much more in case of Lorenz.

\subsection{Hyperchaotic}
Fig.~\ref{14} demonstrates the mTOS effect, $\lambda(4\gamma\rightarrow\infty)<0$, in case of $x_1$-coupled hyperchaotic electronic system. 
Moreover, it is worth noticing that since $\lambda(4\gamma\rightarrow\infty)<0$ implies $\lambda^c<0$ (as shown for R{\"o}ssler), we can say that 
a hyperchaotic drive-response system could also be stabilized without employing BK method~\cite{peng}. Furthermore, similar to 
the above chaotic oscillators, in case of hyperchaotic oscillator feedback scenario yields more robustness but lesser $\beta$ enhancement than the scenario of 
no-feedback and the state of synchronization manifold does not effect the maximization of $\gamma_{dsyn}$ (results not shown). 

\section{Conclusions}
In the present work, without altering the network topology and without employing multi parameter BK method (previous approaches), 
an effort has been made to stabilize the chaotic as well as hyperchaotic synchronization manifold at large coupling strength 
in case of scalar coupling, just by modifying the node dynamics from a single oscillator (SOS) to a pair of oscillators (TOS). 
The impact of modified node configuration for both chaotic and hyperchaotic dynamics, has been studied using Master Stability Function MSF 
(for a fundamental non-centralized ring network) along with the drive-response framework (for a two node network). 
The presented results evidently dictates the potency of TOS in mTOS configuration, i.e. maximization of $\gamma_{dsyn}$ or 
$\lambda(4\gamma\rightarrow\infty)<0$ or $\lambda^c<0$, under the broad range of parametric fluctuations. Furthermore, 
results also demonstrate the enhancement of $\beta$ even when the threshold ($\gamma_{syn}$) augments to large values. Therefore, we believe 
that the proposed TOS would play a vital role both in the fields of complex system as well as secure communication wherein the network 
synchronizability at large coupling strength in case of scalar coupling, remains a primary concern especially for the hyperchaotic systems.

\section{Acknowledgement}
~\newline 
Author would like to thank Vandana H. Singh for her valuable 
suggestions to improve the manuscript presentation. Financial support (501T015013, 502T016522 and TJC23) 
from Saitama University Japan, is acknowledged.


\begin{thebibliography}{50}
\bibitem{qs} If $y$ exhibits steady behavior then for $\gamma_1=\gamma_{1_{01}}$ and $\gamma_2=0$, $\lambda_{0}<0$, 
i.e. the given network should also exhibit steady state behavior ~\cite{monte,russo} where the magnitude of 
$\gamma_{1_{01}}$ depends on the chosen model as $(\gamma_{1_{01}})_{Lorenz}>(\gamma_{1_{01}})_{Chua}>(\gamma_{1_{01}})_{Rossler}$ .
\bibitem{tos} The Lyapunov exponent parallel to synchronization manifold, $\lambda_{0}$, is same for both dTOS and mTOS 
since for $k=0$ mode, dTOS and mTOS follow same equation, i.e. $\dot{\eta}'^0=D\eta'^0$.
\end{thebibliography}
\end{document}